\documentstyle[preprint,aps]{revtex}
\tighten
\begin{document}
\draft
\preprint{\vbox{Submitted to Physical Review C
                        \hfill FSU-SCRI-98-132}}
\title{Lessons to be learned from the coherent \\
       photoproduction of pseudoscalar mesons} 
\author{L.J. Abu-Raddad$^{1,2}$, J. Piekarewicz$^{1,2}$, 
        A.J. Sarty$^{1}$, and R.A.~Rego$^{3}$}
\address{${}^{1}$Department of Physics, 
                 Florida State University, 
                 Tallahassee, FL 32306, USA}
\address{${}^{2}$Supercomputer Computations Research Institute, \\
                 Florida State University, 
	         Tallahassee, FL 32306, USA}
\address{${}^{3}$Instituto de Estudos Avancnados,
                 Centro Tecnico Aeroespacial, \\ 
		 Sao Jose dos Campos, Sao Paulo, Brazil}
\date{\today}
\maketitle
 
\begin{abstract}
We study the coherent photoproduction of pseudoscalar
mesons---particularly of neutral pions---placing special emphasis on
the various sources that put into question earlier
nonrelativistic-impulse-approximation calculations.  These include:
final-state interactions, relativistic effects, off-shell ambiguities,
and violations to the impulse approximation.  We establish that, while
distortions play an essential role in the modification of the coherent
cross section, the uncertainty in our results due to the various
choices of optical-potential models is relatively small (of at most
30\%).  By far the largest uncertainty emerges from the ambiguity in
extending the many on-shell-equivalent representations of the
elementary amplitude off the mass shell.  Indeed, relativistic
impulse-approximation calculations that include the same pionic
distortions, the same nuclear-structure model, and two sets of
elementary amplitudes that are identical on-shell, lead to variations
in the magnitude of the coherent cross section by up to factors of
five.  Finally, we address qualitatively the assumption of locality
implicit in most impulse-approximation treatments, and suggest that
the coherent reaction probes---in addition to the nuclear
density---the polarization structure of the nucleus.
\end{abstract}

\narrowtext

\section{Introduction}
\label{sec:intro}
The coherent photoproduction of pseudoscalar mesons has been
advertised as one of the cleanest probes for studying how
nucleon-resonance formation, propagation, and decay get modified in
the many-body environment; for current experimental efforts see
Ref.~\cite{Sambeek98}. The reason behind such optimism is the
perceived insensitivity of the reaction to nuclear-structure
effects. Indeed, many of the earlier nonrelativistic calculations
suggest that the full nuclear contribution to the coherent process
appears in the form of its matter
density~\cite{bofmir86,cek87,bentan90,tryfik94}---itself believed to
be well constrained from electron-scattering experiments and isospin
considerations.

Recently, however, this simple picture has been put into question.
Among the many issues currently addressed---and to a large extent
ignored in all earlier analyses---are: background (non-resonant)
processes, relativity, off-shell ambiguities, non-localities, and
violations to the impulse approximation. We discuss each one of them
in the manuscript. For example, background contributions to the
resonance-dominated process can contaminate the analysis due to
interference effects.  We have shown this recently for the
$\eta$-photoproduction process, where the background contribution
(generated by $\omega$-meson exchange) is in fact larger than the
corresponding contribution from the $D_{13}(1520)$
resonance~\cite{pisabe97}. In that same study, as in a subsequent
one~\cite{aps98}, we suggested that---by using a relativistic and
model-independent parameterization of the elementary $\gamma
N\rightarrow \eta N$ amplitude---the nuclear-structure information
becomes sensitive to off-shell ambiguities. Further, the local
assumption implicit in most impulse-approximation calculations, and
used to establish that all nuclear-structure effects appear
exclusively via the matter density, has been lifted by Peters, Lenske,
and Mosel~\cite{Peters98}. An interesting result that emerges from
their work on coherent $\eta$-photoproduction is that the
$S_{11}(1535)$ resonance---known to be dominant in the elementary
process but predicted to be absent from the coherent
reaction~\cite{bentan90}---appears to make a non-negligible
contribution to the coherent process. Finally, to our knowledge, a
comprehensive study of possible violations to the
impulse-approximation, such as the modification to the production,
propagation, and decay of nucleon resonances in the nuclear medium,
has yet to be done.

In this paper we concentrate---in part because of the expected
abundance of new, high-quality experimental data---on the 
coherent photoproduction of neutral pions. The central issue 
to be addressed here is the off-shell ambiguity that emerges 
in relativistic descriptions and its impact on extracting 
reliable resonance parameters; no attempt has been made here 
to study possible violations to the impulse approximation or 
to the local assumption. Indeed, we carry out our calculations 
within the framework of a relativistic impulse approximation 
model. However, rather than resorting to a nonrelativistic
reduction of the elementary $\gamma N\rightarrow\pi^{0}N$ amplitude,
we keep intact its full relativistic structure~\cite{cgln57}. As a
result, the lower components of the in-medium Dirac spinors are
evaluated dynamically in the Walecka model~\cite{serwal86}. 

Another important ingredient of the calculation is the final-state
interactions of the outgoing pion with the nucleus. We address the
pionic distortions via an optical-potential model of the pion-nucleus
interaction. We use earlier models of the pion-nucleus interaction
plus isospin symmetry---since these models are constrained mostly from
charged-pion data---to construct the neutral-pion optical potential.
However, since we are unaware of a realistic optical-potential model
that covers the $\Delta$-resonance region, we have extended the
low-energy work of Carr, Stricker-Bauer, and McManus~\cite{SMC} to
higher energies. In this way we have attempted to keep at a minimum
the uncertainties arising from the optical potential, allowing
concentration on the impact of the off-shell ambiguities to the coherent
process. A paper discussing this extended optical-potential model will
be presented shortly~\cite{raddad98}. Finally, we use an elementary
$\gamma N\rightarrow\pi^{0}N$ amplitude extracted from the most recent
phase-shift analysis of Arndt, Strakovsky, and Workman~\cite{Arnphn}.

Our paper has been organized as follows. In Sec.~\ref{sec:distor} and
in the appendix we discuss in some detail the pion-nucleus interaction
and its extension to the $\Delta$-resonance region.
Sec.~\ref{sec:off-shell} is devoted to the central topic of the paper:
the large impact of the off-shell ambiguity on the coherent cross
section. Sec.~\ref{sec:impviol} includes a qualitative discussion on
several important mechanisms that go beyond the impulse-approximation
framework, but that should, nevertheless, be included in any proper
treatment of the coherent process.  Finally, we summarize in
Sec.~\ref{sec:concl}.

\section{Pionic Distortions}
\label{sec:distor}
	Pionic distortions play a critical role in all studies
	involving pion-nucleus interactions. These distortions are
	strong and, thus, modify significantly any process relative to
	its naive plane-wave limit. Indeed, it has been shown in
	earlier studies of the coherent pion photoproduction
	process---and verified experimentally~\cite{ndu91}---that
	there is a large modification of the plane-wave cross section
	once distortions are included.  Because of the importance of
	the pionic distortions, any realistic study of the coherent
	reaction must invoke them from the outset. However, since a
	detailed microscopic model for the distortions has yet to be
	developed, we have resorted to an optical-potential
	model. This semi-phenomenological choice implies some
	uncertainties. Thus, pionic distortions represent the first
	challenge in dealing with the coherent photoproduction
	processes.

        We have used earlier optical-potential models of the
	pion-nucleus interaction, supplemented by isospin symmetry, to
	construct the $\pi^0$-nucleus optical potential. Moreover, we
	have extended the low-energy work of Carr, Stricker-Bauer, and
	McManus~\cite{SMC} to the $\Delta$-resonance region.  Most of
	the formal aspects of the optical potential have been reserved
	to the appendix and to a forthcoming
	publication~\cite{raddad98}.  Here we proceed directly to
	discuss the impact of the various choices of optical
	potentials on the coherent cross section.

\subsection{Results}
\label{subsec:distresults}

\begin{figure}[h]
 \null
 \vskip1.2in
 \includegraphics{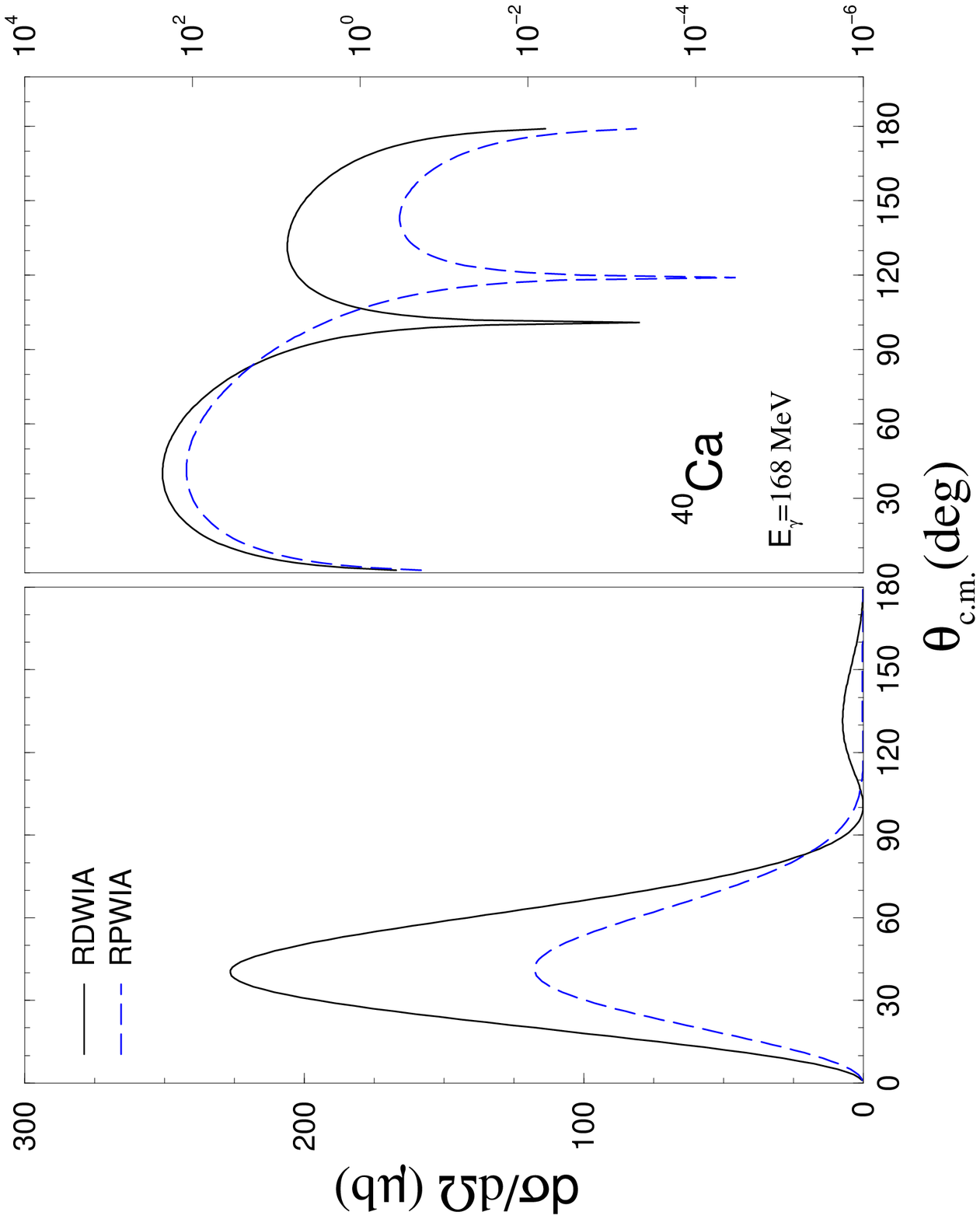}
 \vskip2.0in
\caption{Differential cross section for the coherent pion 
         photoproduction reaction from $^{40}\rm{Ca}$ at
         $E_{\gamma}=168$ MeV using the vector representation
         for the elementary amplitude with (solid line) and
         without (dashed line) the inclusion of distortions.
	 Results on the left(right) panel are plotted using
         a linear(logarithmic) scale.}
\label{fig1}
\end{figure}

	The large effect of distortions can be easily seen in
	Fig.~\ref{fig1}. The left panel of the graph (plotted on a
	linear scale) shows the differential cross section for the
	coherent photoproduction of neutral pions from $^{40}\rm{Ca}$
	at a laboratory energy of $E_{\gamma}\!=\!168$~MeV. The solid
	line displays our results using a relativistic distorted-wave
	impulse approximation (RDWIA) formalism, while the dashed line
	displays the corresponding plane-wave result (RPWIA). The
	calculations have been done using a vector representation for
	the elementary $\gamma N \rightarrow \pi^{0} N$
	amplitude. Note that this is only one of the many possible
	representations of the elementary amplitude that are
	equivalent on-shell. A detailed discussion of these off-shell
	ambiguities is deferred to Sec.~\ref{sec:off-shell}.
	At this specific photon energy---one not very far 
	from threshold---the distortions have more than 
	doubled the value of the differential cross section 
	at its maximum. Yet, the shape of the angular 
	distribution seems to be preserved. However, upon
	closer examination (the right panel of the graph 
	shows the same calculations on a logarithmic scale)
	we observe that the distortions have caused a 
	substantial back-angle enhancement due to a 
	different sampling of the nuclear density, relative 
	to the plane-wave calculation. This has resulted in 
	a small---but not negligible---shift of about
 	$10^{\circ}$ in the position of the minima. The
	back-angle enhancement, with its corresponding 
	shift in the position of the minimum, has been 
	seen in our calculations also at different incident 
        photon energies.

\begin{figure}[h]
 \null
 \vskip1.2in
 \includegraphics{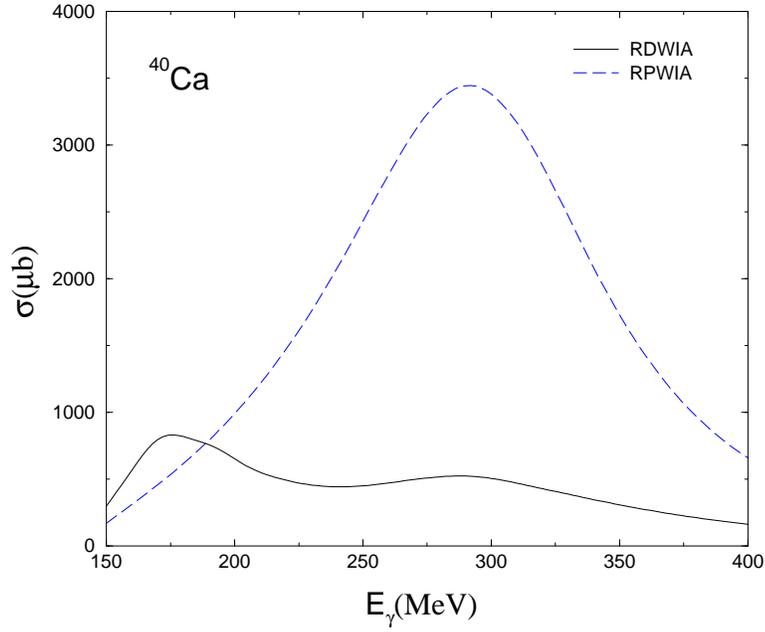}
 \vskip2.0in
\caption{Total cross section for the coherent pion 
         photoproduction reaction from $^{40}\rm{Ca}$ as
	 a function of the photon energy in the laboratory
	 frame with (solid line) and without (dashed line)
	 including pionic distortions. A vector
	 representation for the elementary part of the
	 amplitude is used.}
\label{fig2}
\end{figure}

The effect of distortions on the total photoproduction  cross section
from $^{40}\rm{Ca}$ as a function of the photon energy is displayed in
Fig.~\ref{fig2}. The behavior of the the distorted cross section is
explained in terms of a competition between the attractive real
(dispersive) part and the absorptive imaginary part of the optical
potential. Although the optical potential encompasses very complicated
processes, the essence of the physics can be understood in terms of
$\Delta$-resonance dominance.  Ironically, the behavior of the
dispersive and the absorptive parts are caused primarily by the same
mechanism: $\Delta$-resonance formation in the nucleus. The mechanism
behind the attractive real part is the scattering of the pion
from a single nucleon---which is dramatically increased in the
$\Delta$-resonance region. In contrast, the absorptive imaginary part
is the result of several mechanisms, such as nucleon knock-out,
excitation of nuclear states, and two-nucleon processes. At very low
energies some of the absorptive channels are not open yet, resulting
in a small imaginary part of the potential. This in turn provides a
chance for the attractive real part to enhance the coherent cross
section. As the energy increases, specifically in the
$\Delta$-resonance region, a larger number of absorptive channels
become available leading to a large dampening of the cross section.
Although the attractive part also increases around the
$\Delta$-resonance region, this increase is more than compensated by
the absorptive part, which greatly reduces the probability for the 
pion to interact elastically with the nucleus.

	Since understanding pionic distortions constitutes our first
	step towards a comprehensive study of the coherent process, it
	is instructive to examine the sensitivity of our results to
	various theoretical models. To this end, we have calculated
	the coherent cross section using different optical potentials,
	all of which fit $\pi$-nucleus scattering data as well as the
	properties of pionic atoms. We have started by calculating
	the coherent cross section using the optical potential
	developed by Carr and collaborators~\cite{SMC}.  It should be
	noted that although our optical potential originates from the
	work of Carr and collaborators, there are still significant
	differences between the two sets of optical potentials. Some
	of these differences arise in the manner in which some
	parameters are determined. Indeed, in our case parameters
	that have their origin in pion--single-nucleon physics
	have been determined from a recent $\pi\!-\!N$
	phase shift analysis~\cite{Arnpin}, while Carr and
	collaborators have determined them from fits to pionic-atom
	data. Moreover, we have included effects that were not
	explicitly included in their model, such as Coulomb
	corrections when fitting to charge-pion data. 

	In addition to the above potentials, we have calculated the 
	coherent cross section using a simple 4-parameter Kisslinger 
	potential of the form:
\begin{equation}
	2 \omega U \!=\!-4\pi 
        \left[
	 b_{\rm eff} \rho(r) - c_{\rm eff}\vec{\nabla}
	 \rho(r)\cdot \vec{\nabla} + c_{\rm eff} 
	 \frac{\omega}{2 M_N} {\nabla}^2\rho(r)
        \right].
\end{equation} 
Note that we have used two different sets of parameters for
this Kisslinger potential, denoted by K1 and K2~\cite{SMC}. 
Both sets of parameters were constrained by $\pi$-nucleus 
scattering data and by the properties of pionic atoms. 
However, while the K1 fit was constrained to obtain 
$b_{\rm eff}$ and $c_{\rm eff}$ parameters that did not 
deviate much from their pionic-atom values, the K2 fit 
allowed them to vary freely, so as to obtain the best 
possible fit. 

\begin{figure}[h]
 \null
 \vskip1.2in
 \includegraphics{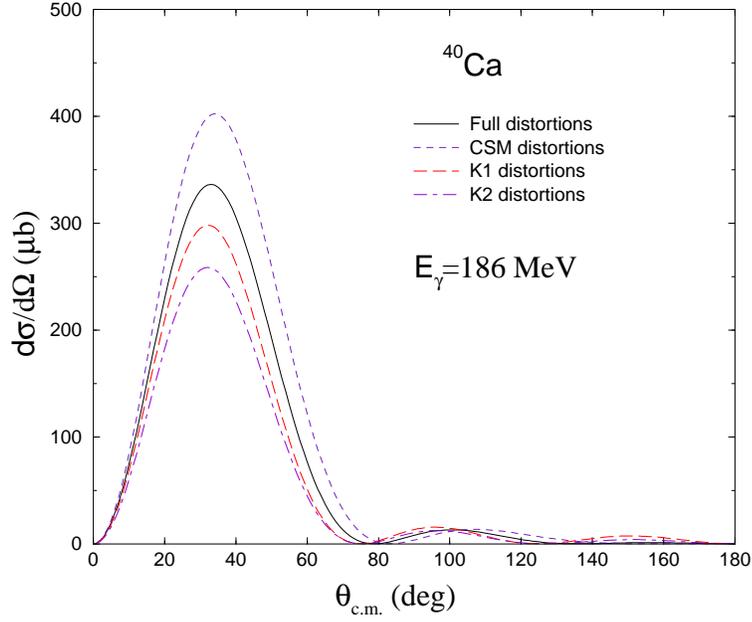}
 \vskip2.0in
\caption{Differential cross section for the coherent 
	 pion-photoproduction reaction from 
	 $^{40}\rm{Ca}$ at 
	 $E_{\gamma}\!=\!186$~MeV 
         (resulting in the emission of a 50~MeV pion)
	 using different 
	 optical-potential models. All of these models are
	 equivalent insofar as they fit 
	 properties of pionic atoms and $\pi$-nucleus 
	 scattering data. A vector representation for the 
	 elementary part of the amplitude is used.}
\label{fig3}
\end{figure}

Results for the coherent photoproduction cross section from
$^{40}\rm{Ca}$ at a photon energy of $E_{\gamma}\!=\!186$~MeV
(resulting in the emission of a 50~MeV pion) for the various
optical-potential models are shown in Fig.~\ref{fig3}.  In the plot,
our results are labeled full-distortions (solid line) while those of
Carr, Stricker-Bauer, and McManus as CSM (short dashed line); those
obtained with the 4-parameter Kisslinger potential are labeled K1
(long-dashed line) and K2 (dot-dashed line), respectively. It can be
seen from the figure that our calculation differs by at most 30\%
relative to the ones using earlier forms of the optical
potential. Note that we have only presented results computed using the
vector parameterization of the elementary amplitude. Similar
calculations done with the tensor amplitude (not shown) display
optical-model uncertainties far smaller (of the order of 5\%) than the
ones reported in Fig.~\ref{fig3}. In conclusion, although there seems
to be a non-negligible uncertainty arising from the optical potential,
these uncertainties pale in comparison to the large off-shell
ambiguity, to be discussed next.

\section{Off-Shell Ambiguity}
\label{sec:off-shell}

	The study of the coherent reaction represents a challenging
	theoretical task due to the lack of a detailed microscopic
	model of the process. Indeed, most of the models used to date
	rely on the impulse approximation: the assumption that the
	elementary $\gamma N \rightarrow \pi N$ amplitude remains
	unchanged as the process is embedded in the nuclear
	medium. Yet, even a detailed knowledge of the elementary
	amplitude does not guarantee a good understanding of the
	coherent process.  The main difficulty stems from the fact
	that there are, literally, an infinite number of equivalent
	on-shell representations of the elementary amplitude.  These
	different representations of the elementary
	amplitude---although equivalent on-shell---can give very
	different results when evaluated off-shell. Of course, this
	uncertainty is present in many other kind of nuclear
	reactions, not just in the coherent photoproduction process.
	Yet, this off-shell ambiguity comprises one of the
	biggest, if not the biggest, hurdle in understanding the
	coherent photoproduction of pseudoscalar mesons.

\subsection{Formalism}
\label{subsec:offformal}

	Before discussing the off-shell ambiguity, let us set
        the background by introducing some model-independent
	results for the differential cross section. Using the
	relativistic formalism developed in our earlier 
	work~\cite{pisabe97}, the differential cross
	section in the center-of-momentum frame (c.m.)
	for the coherent photoproduction of pseudoscalar
	mesons is given by
\begin{equation}
   \left({d\sigma \over d\Omega}\right)_{\rm c.m.}=
    \left({M_{\lower 1pt \hbox{$\scriptstyle T$}} 
    \over 4\pi W}\right)^{2} 
    \left({q_{\rm c.m.} \over k_{\rm c.m.}}\right)
    \left({1 \over 2}k_{\rm c.m.}^{2}q_{\rm c.m.}^{2}
    \sin^{2}\theta_{\rm c.m.}\right)\,
    |F_{0}(s,t)|^{2} \;,
 \label{dsigmab}
\end{equation}
where
	$M_{\lower 1pt \hbox{$\scriptstyle T$}}$ is the mass of the
	target nucleus. Note that $W$, $\theta_{\rm c.m.}$, $k_{\rm
	c.m.}$ and $q_{\rm c.m.}$ are the total energy, scattering
	angle, photon and $\pi$-meson momenta in the c.m. frame,
	respectively. Thus, all dynamical information
	about the coherent process is contained in the single
	Lorentz-invariant form factor $F_{0}(s,t)$; this form-factor
	depends on the Mandelstam variables $s$ and $t$.

 	We now proceed to compute the Lorentz invariant form factor
	in a relativistic impulse approximation. In order to do so, 
	we need an expression for the amplitude of the elementary
        process: $\gamma N \rightarrow \pi^0 N$. We start by
	using the ``standard'' model-independent 
	parameterization given in terms of four Lorentz- and 
	gauge-invariant amplitudes~\cite{bentan90,cgln57}. That is,
\begin{equation}
   T(\gamma N \rightarrow \pi^0 N) =
   \sum_{i=1}^{4} A_{i}(s,t) {M}_{i} \;,
 \label{telema}
\end{equation}
where the $A_{i}(s,t)$ are scalar functions of $s$ and $t$ and
for the Lorentz structure of the amplitude we use the standard 
set:
\begin{mathletters}
 \begin{eqnarray}
  {M}_{1} &=& -\gamma^{5}\rlap/{\epsilon}\;\rlap/{k} \;, \\
  {M}_{2} &=& 2\gamma^{5}
    \Big[(\epsilon \cdot p)(k \cdot p') -
         (\epsilon \cdot p')(k \cdot p) \Big]  \;, \\
  {M}_{3} &=& \gamma^{5}
    \Big[\rlap/{\epsilon}\,(k \cdot p) -
         \rlap/{k}(\epsilon \cdot p) \Big]     \;, \\
  {M}_{4} &=& \gamma^{5}
    \Big[\rlap/{\epsilon}\,(k \cdot p') -
         \rlap/{k}(\epsilon \cdot p') \Big]    \;. 
  \label{allm}
 \end{eqnarray}
\end{mathletters}
This form, although standard, is only one particular choice
for the elementary amplitude. Many other choices---all of them 
equivalent on shell---are possible. Indeed, we could have used 
the relation---valid only on the mass shell,
 \begin{eqnarray}
  {M}_{1} &=& -\gamma^{5}\rlap/{\epsilon}\;\rlap/{k} 
           =  {1\over 2}
   	      \varepsilon^{\mu\nu\alpha\beta}\,
              \epsilon_{\mu}\,k_{\nu} \sigma_{\alpha \beta}            
           =  {i\over 2}
   	      \varepsilon^{\mu\nu\alpha\beta}\,
              \epsilon_{\mu}\,k_{\nu}\,{Q_{\alpha}\over M_N}\gamma_{\beta}
              \nonumber \\
          &-&  {1 \over 2M_N}\gamma^{5}
              \Big[\rlap/{\epsilon}\,(k \cdot p) -
                   \rlap/{k}(\epsilon \cdot p) \Big] 
           -  {1 \over 2M_N}\gamma^{5}
              \Big[\rlap/{\epsilon}\,(k \cdot p') -
              \rlap/{k}(\epsilon \cdot p') \Big] \;,
 \end{eqnarray}
to obtain the following representation of the elementary amplitude:
\begin{equation}
   T(\gamma N \rightarrow \pi^0 N) =
   \sum_{i=1}^{4} B_{i}(s,t) {N}_{i} \;.
 \label{telemavec}
\end{equation}
where the new invariant amplitudes and Lorentz structures
are now defined as:
\begin{mathletters}
 \begin{eqnarray}
    B_{1} &=& A_{1}          \;; \qquad \phantom{-,A_{1}/2M} \;\;
    N_{1}  = {i\over 2}
    \varepsilon^{\mu\nu\alpha\beta}\,
    \epsilon_{\mu}\,k_{\nu}\,{Q_{\alpha}\over M_N}\gamma_{\beta} \;, \\
    B_{2} &=& A_{2}          \;; \qquad \phantom{-,A_{1}/2M_N}
    N_{2}  = M_{2}=2\gamma^{5}
    \Big[(\epsilon \cdot p)(k \cdot p') -
         (\epsilon \cdot p')(k \cdot p) \Big]  \;, \\
    B_{3} &=& A_{3}-A_{1}/2M_N \;; \qquad 
    N_{3}  = M_{3}=\gamma^{5}
    \Big[\rlap/{\epsilon}\,(k \cdot p) -
         \rlap/{k}(\epsilon \cdot p) \Big]     \;, \\
    B_{4} &=& A_{4}-A_{1}/2M_N \;; \qquad
    N_{4}  = M_{4}=\gamma^{5}
    \Big[\rlap/{\epsilon}\,(k \cdot p') -
         \rlap/{k}(\epsilon \cdot p') \Big]    \;. 
 \end{eqnarray}
\end{mathletters}
Note that we have introduced the four-momentum transfer
$Q^{\mu}\!\equiv\!(k-q)^{\mu}\!=\!(p'-p)^{\mu}$. Although clearly
different, Eqs.~(\ref{telema}) and (\ref{telemavec}) are totally
equivalent on-shell: no observable measured in the elementary $\gamma
N \rightarrow \pi^0 N$ process could distinguish between these two
forms. We could go on. Indeed, it is well known that a pseudoscalar
and a pseudovector representation are equivalent on shell. That is, we
could substitute the pseudoscalar vertex in $N_{2}$ and $M_{2}$ by a
pseudovector one:
\begin{equation}
  \gamma^{5}={\rlap/{\!Q}\over 2M_N}\gamma^{5} \;.
\end{equation}
The possibilities seem endless.

	Given the fact that there are many---indeed
	infinite---equivalent parameterizations of  the elementary
	amplitude on-shell, it becomes  ambiguous on how to take the
	amplitude off the mass shell.  In this work we have examined
	this off-shell ambiguity  by studying the coherent process
	using the ``tensor'' parameterization, as in
	Eq.~(\ref{telema}), and the  ``vector'' parameterization, as
	in Eq.~(\ref{telemavec}). Denoting these parameterizations as
	tensor and vector originates from the fact that for the
	coherent process from spherical nuclei (such as the ones
	considered here) the respective cross sections become
	sensitive to only the tensor and vector densities,
	respectively. Indeed, the tensor parameterization yields a
	coherent amplitude  that depends exclusively on the	
	ground-state tensor density~\cite{pisabe97}:
\begin{equation}
  \Big[
    \rho_{\lower 3pt \hbox{$\scriptstyle T$}}(r)\,\hat{r}
  \Big]^{i} = 
  \sum_{\alpha}^{\rm occ}
   \overline{{\cal U}}_{\alpha}({\bf x})\,
   \sigma^{{\scriptscriptstyle 0}i} \,
              {\cal U}_{\alpha}({\bf x}) \;; \quad
   \rho_{\lower 3pt \hbox{$\scriptstyle T$}}(r) =
   \sum_{a}^{\rm occ}
   \left({2j_{a}+1 \over 4\pi r^{2}}\right)
   2g_{a}(r)f_{a}(r) \;,
 \label{rhotr}
 \end{equation}
where ${\cal U}_{\alpha}({\bf x})$ is an in-medium
single-particle Dirac spinor, $g_{a}(r)$ and $f_{a}(r)$ 
are the radial parts of the upper and lower components of 
the Dirac spinor, respectively, and  the above sums
run over all the occupied single-particle states.

	The vector parameterization, on the other hand, leads 
to a coherent amplitude that depends on timelike-vector---or 
matter---density of the nucleus which is defined as:
\begin{equation}
  \rho_{\lower 3pt \hbox{$\scriptstyle V$}}(r) =
  \sum_{\alpha}^{\rm occ}
  \overline{{\cal U}}_{\alpha}({\bf x})\,
  \gamma^{{\scriptscriptstyle 0}} \,
         {\cal U}_{\alpha}({\bf x}) \;; \quad
   \rho_{\lower 3pt \hbox{$\scriptstyle V$}}(r) =
   \sum_{a}^{\rm occ}
   \left({2j_{a}+1 \over 4\pi r^{2}}\right)
   \Big(g_{a}^2(r) + f_{a}^2(r)\Big) \;.
 \label{rhov}
\end{equation}
In determining these relativistic ground-state densities, we 
have used a mean-field approximation to the 
Walecka model~\cite{serwal86}. In doing so, we have maintained 
the full relativistic structure of the process. In the Walecka
model, one obtains three non-vanishing ground state densities 
for spherical, spin-saturated nuclei. These are the timelike-vector
and tensor densities defined earlier, and the scalar density given 
by 
\begin{equation}
  \rho_{\lower 3pt \hbox{$\scriptstyle S$}}(r) = 
  \sum_{\alpha}^{\rm occ}
   \overline{{\cal U}}_{\alpha}({\bf x})\,
             {\cal U}_{\alpha}({\bf x}) \;; \quad
   \rho_{\lower 3pt \hbox{$\scriptstyle S$}}(r) =
   \sum_{a}^{\rm occ}
   \left({2j_{a}+1 \over 4\pi r^{2}}\right)
   \Big(g_{a}^2(r) - f_{a}^2(r)\Big) \;.
 \label{rhos}
\end{equation}
All other ground-state densities---such as the pseudoscalar and
axial-vector densities---vanish due to parity conservation.  This is
one of the appealing features of the coherent reaction; because of the
conservation of parity, the coherent process becomes sensitive to only
one ($A_{1}$) of the possible four, elementary amplitudes. It is
important to note that the three non-vanishing relativistic
ground-state densities are truly independent and constitute
fundamental nuclear-structure quantities.  The fact that in the
nonrelativistic
framework~\cite{bofmir86,cek87,bentan90,tryfik94,ndu91} only one
density survives (the scalar and vector densities become equal and the
tensor density becomes dependent on the vector one) is due to the
limitation of the approach.  Indeed, in the nonrelativistic framework
one employs free Dirac spinors to carry out the nonrelativistic
reduction of the elementary amplitude. Hence, any evidence of possible
medium modifications to the ratio of lower-to-upper components of the
Dirac spinors is lost.

Before presenting our results we should mention a ``conventional''
off-shell ambiguity. In the vector parameterization of
Eq.~(\ref{telemavec}) the amplitude includes the four momentum
transfer $Q\!=\!(k\!-\!q)$. While the photon momentum ${\bf k}$ 
is well defined, the asymptotic pion three-momentum ${\bf q}$
is different---because of distortions---from the pion momentum 
immediately after the photoproduction process. Since the ``local''
pion momentum in the interaction region is the physically relevant 
quantity, we have replaced the asymptotic pion momentum ${\bf q}$ 
by the pion-momentum operator ($-i{\bf \nabla}$). Thus, 
in evaluating the scattering matrix element $T_{\pm}\!=\!\langle 
\pi(q);A(p')| J^{\mu} |A(p);\gamma(k,\epsilon_{\pm})\rangle$, 
we arrive at an integral of the form:
\begin{eqnarray}	
  \varepsilon^{i j m 0} \epsilon_i k_j
  \large \int d^{3}x 
  \Big[
    {{\bf\nabla}}{{\phi}^{(-)}_{q}}^{*}({\bf x})
  \Big]_m 
  e^{i{\bf k}\cdot{\bf x}}
  \frac{\rho_{\lower 3pt\hbox{$\scriptscriptstyle V$}}(r)}{2M_N}
  &=&\pm (2\pi)^{3/2}\frac{|\bf k|}{M_N} 
  \sum_{l=1}^{\infty}
  \sqrt{\frac{l(l+1)}{2l+1}} \nonumber\\ 
  & & {Y}_{l,\pm 1}(\hat{q}) \int r^2 dr 
  \rho_{\lower 3pt\hbox{$\scriptstyle V$}}(r) R_{l}(r),
\end{eqnarray}
where
\begin{equation}
    R_{l}(r) =  
    j_{l+1}(kr)\left[\frac{d}{dr}-\frac{l}{r}\right]
    {\phi}^{(+)}_{l,q}(r) +
    j_{l-1}(kr)\left[\frac{d}{dr}+\frac{l\!+\!1}{r}\right]
    {\phi}^{(+)}_{l,q}(r)\;.
\end{equation}
Note that we have introduced the distorted pion wave function
${\phi}^{(\pm)}_{q}({\bf x})$, the spherical Bessel functions of order
${l\pm 1}$, and the $\pm$ sign for positive/negative circular
polarization of the incident photon. Moreover, adopting the ${\bf q}
\rightarrow -i{\bf \nabla}$ prescription, has resulted, as in the
tensor case \cite{pisabe97}, in no s-wave ($l\!=\!0$) contribution to
the scattering amplitude. This is also in agreement with the earlier
nonrelativistic calculation of Ref.~\cite{bofmir86}. Finally, we have
obtained the four Lorentz- and gauge-invariant amplitudes $A_{i}(s,t)$
for the elementary process from the phase-shift analysis of the VPI
group~\cite{Arnphn}.

\subsection{Results}
\label{subsec:offresults}
\begin{figure}[h]
 \null
 \vskip1.2in
 \includegraphics{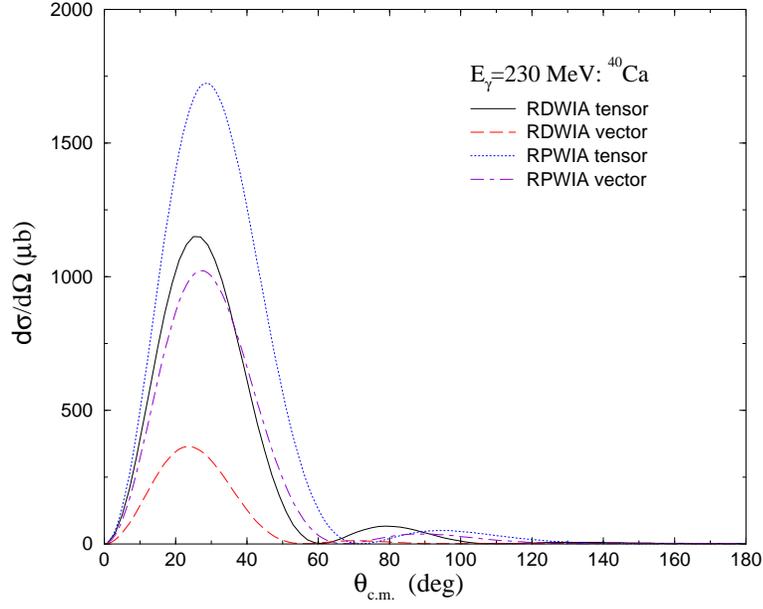}
 \vskip2.0in
\caption{Differential cross section for the coherent pion 
	 photoproduction reaction from $^{40}\rm{Ca}$ at
	 $E_{\gamma}\!=\!230$ MeV with (RDWIA) and without 
	 (RPWIA) pionic distortions. Tensor and vector
	 parameterizations of the elementary amplitude 
	 are used.}
\label{fig4}
\end{figure}

Based on the above formalism, we present in Fig.~\ref{fig4} the
differential cross section for the coherent photoproduction of neutral
pions from $^{40}\rm{Ca}$ at a photon energy of $E_{\gamma}\!=\!230$
MeV using a relativistic impulse approximation approach. Both tensor
and vector parameterizations of the elementary amplitude were
used. The off-shell ambiguity is immense; factors of two (or more) are
observed when comparing the vector and tensor representations. It is
important to stress that these calculations were done by using the
same nuclear-structure model, the same pionic distortions, and two
elementary amplitudes that are identical on-shell. The very large
discrepancy between the two theoretical models emerges from the
dynamical modification of the Dirac spinors in the nuclear
medium. Indeed, in the nuclear medium the tensor density---which  is
linear in the lower-component of the Dirac spinors [see
Eq.~(\ref{rhotr})]---is strongly enhanced due to the presence of  a
large scalar potential (the so-called ``$M^{\star}$-effect'').  In
contrast, the conserved vector density is insensitive to the
$M^{\star}$-effect. Yet the presence of the large scalar---and
vector---potentials in the nuclear medium is essential in  accounting
for the bulk properties of nuclear matter and finite
nuclei~\cite{serwal86}. We have compared our theoretical results  to
preliminary  and unpublished data (not shown) provided to us  courtesy
of B.~Krusche~\cite{krusche98}. The data follows the  same shape as
our calculations but the experimental curve seems to  straddle between
the two calculations, although the vector calculations appears closer
to the experimental data. This behavior---a closer agreement of the
vector calculation to  data---has been observed in all of the
comparisons that we have done so far.

\begin{figure}[h]
 \null
 \vskip1.2in
 \includegraphics{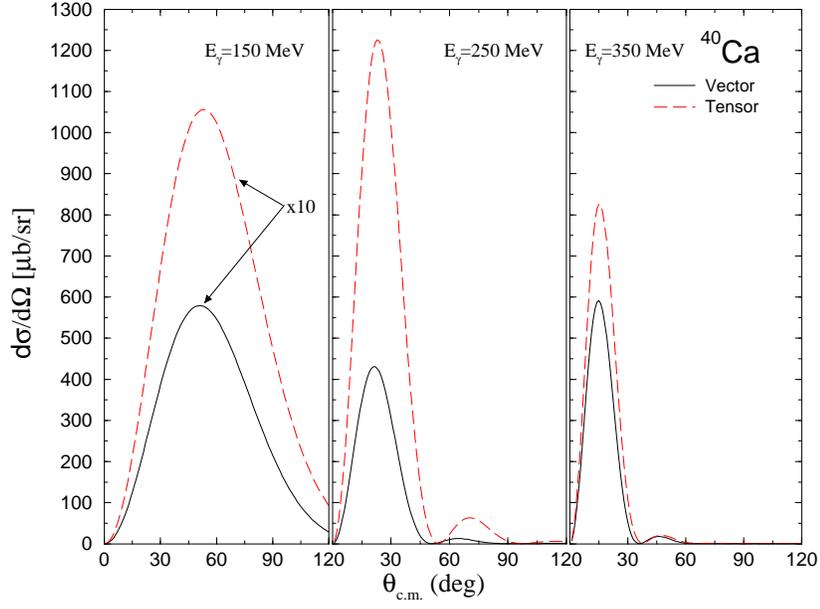}
 \vskip2.0in
\caption{Differential cross section for the coherent pion 
	 photoproduction reaction for $^{40}\rm{Ca}$ at a
	 variety of photon energies using a RDWIA formalism.
	 Tensor (dashed line) and vector (solid line) 
	 parameterizations of the elementary amplitude are 
	 used.} 
\label{fig5}
\end{figure}

\begin{figure}[h]
 \null
 \vskip1.2in
 \includegraphics{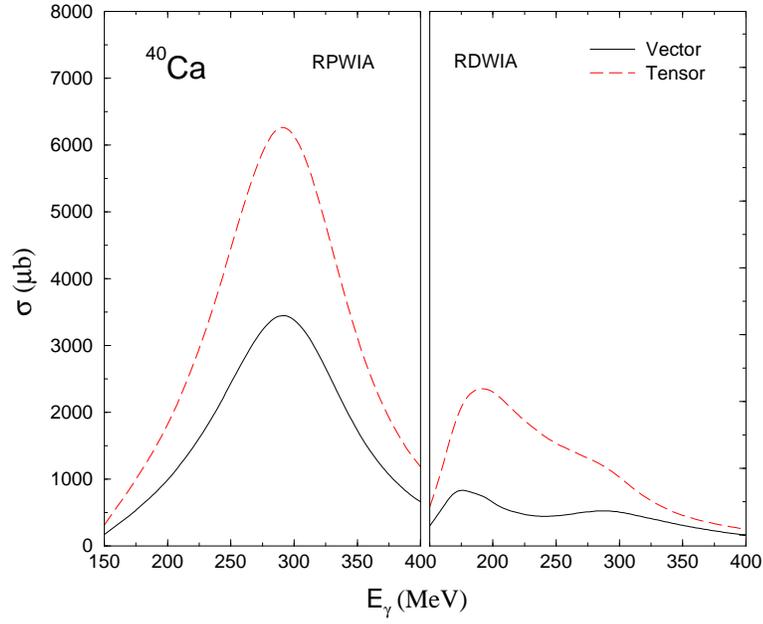}
 \vskip2.0in
\caption{Total cross section for the coherent pion 
	 photoproduction reaction from $^{40}\rm{Ca}$ as a 
	 function of the photon energy with (right panel) and
         without (left panel) pionic distortions.
	 Tensor (dashed line) and vector (solid line) 
	 parameterizations of the elementary amplitude are used.} 
\label{fig6}
\end{figure}

In Fig.~\ref{fig5} we present results for the differential 
cross section from $^{40}\rm{Ca}$ at a variety of photon energies, 
while in Fig.~\ref{fig6} we display results for the total cross 
section. By examining these graphs one can infer that the tensor 
parameterization always predicts a large enhancement of the cross
section---irrespective of the photon incident energy and the
scattering angle---relative to the vector predictions. As stated
earlier, this large enhancement is inextricably linked to the
corresponding in-medium enhancement of the lower components of 
the nucleon spinors. Moreover, the convolution of the tensor and 
vector densities with the pionic distortions give rise to similar 
qualitative, but quite different quantitative, behavior on the energy 
dependence of the corresponding coherent cross sections.

\begin{figure}[h]
 \null
 \vskip1.2in
 \includegraphics{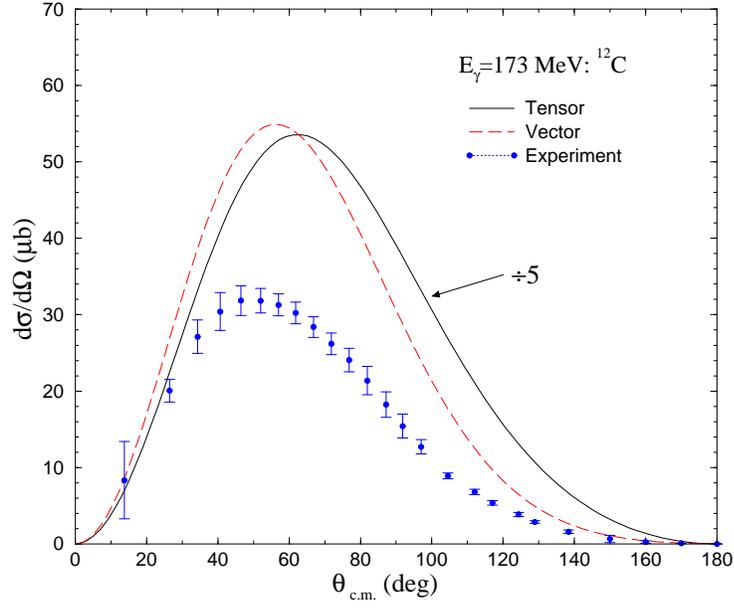}
 \vskip2.0in
\caption{Differential cross section for the coherent pion
	 photoproduction reaction from $^{12}\rm{C}$ 
	 at $E_{\gamma}\!=\!173$ MeV. Tensor (dashed line) 
	 and vector (solid line) parameterizations of the 
	 elementary amplitude are used. The experimental 
	 data is from Ref.~\protect\cite{gothe95}.} 
\label{fig7}
\end{figure}

In order to explore the A-dependence of the coherent process, we have
also calculated the cross section from $^{12}$C at various photon
energies. This is particularly relevant for our present discussion, as
$^{12}$C displays an even larger off-shell ambiguity than
$^{40}$Ca. In Fig.~\ref{fig7} we show the differential cross section
for the coherent process from $^{12}$C at a photon energy of
$E_{\gamma}\!=\!173$~MeV. The off-shell ambiguity for this case is
striking; at this energy the tensor result is five times larger than
the vector prediction.  The additional enhancement observed here
relative to $^{40}$Ca is easy to understand on the basis of some of
our earlier work~\cite{aps98}. Indeed, we have shown in our study of
the coherent photoproduction of $\eta$-mesons, that if one
artificially adopts an in-medium ratio of upper-to-lower components
identical to the one in free space, then the tensor and vector
densities are no longer independent; rather, they become related by:
\begin{equation}
   \rho_{\lower 3pt \hbox{$\scriptstyle T$}}(Q) 
   = - \frac{Q}{2M_{N}}
   \rho_{\lower 3pt \hbox{$\scriptstyle V$}}(Q) \;. 
 \label{freefg}
\end{equation}   
However, this relation was proven to be valid only for closed-shell 
nuclei. As ${}^{12}$C is an open-shell nucleus (closed $p^{3/2}$
but open $p^{1/2}$ orbitals) an additional enhancement of the tensor 
density---above and beyond the $M^{\star}$-effect---was observed.
Fig.~\ref{fig7} also shows a comparison of our results with the
experimental data of Ref.~\cite{gothe95}. It is clear from the
figure that the vector representation is closer to the data; note
that the tensor calculation has been divided by a factor of five. 
Even so, the vector calculation also overestimates the data by a
considerable amount.

For further comparison with experimental data we have calculated the
coherent cross section from ${}^{12}$C at photon energies of
$E_{\gamma}\!=\!235$, $250$, and $291$~MeV. In Table~\ref{table1} we
have collated our calculations with experimental data published by
J. Arends and collaborators~\cite{Arends83} for $E_{\gamma}\!=\! 235$
and $291$ MeV, and with data presented by Booth~\cite{Booth1978} and
A. Nagl, V. Devanathan, and H. \"Uberall~\cite{ndu91} for $E_{\gamma
\rm{lab}}\!=\!250$~MeV.  The experimental data exhibits similar
patterns as our calculations (not shown) but the values of the maxima
of the cross section are different. The tensor calculations continue
to predict large enhancement factors (of five and more) relative to
the vector calculations. More importantly, these enhancement
factors are in contradiction with experiment. The experimental data
appears to indicate that the maximum in the differential cross section
from $^{12}\rm{C}$ is largest at about $250$ MeV, while our
calculations predict a maximum around $295$ MeV.
It is likely that this energy ``shift'' might be the result of the 
formation and propagation of the $\Delta$-resonance in the nuclear
medium. Clearly, in an impulse-approximation framework, medium
modifications to the elementary amplitude---arising from changes in
resonance properties---can not be accounted for. Yet, a 
binding-energy correction of about $40$~MeV due to the 
$\Delta$-nucleus interaction has been suggested before. Indeed, such
a shift would also explain the discrepancy in the position of our
theoretical cross sections in $^{40}$Ca, relative to the
(unpublished) data by Krusche and collaborators~\cite{krusche98}.
Moreover, such a shift---albeit of only 15 MeV---was invoked by
Peters, Lenske, and Mosel~\cite{Peters98} in their recent calculation
of the coherent pion-photoproduction cross section. Yet, a detailed
study of modifications to hadronic properties in the nuclear medium
must go beyond the impulse approximation; a topic outside the
scope of the present work. However, a brief qualitative discussion 
of possible violations to the impulse approximation is given in the
next section.

\begin{figure}[h]
 \null
 \vskip1.2in
 \includegraphics{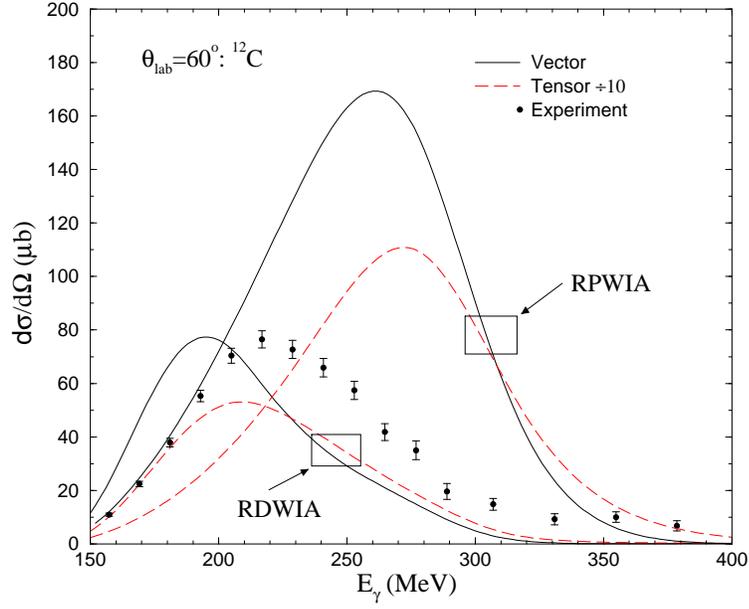}
 \vskip2.0in
\caption{Differential cross section for the coherent pion
	 photoproduction reaction from $^{12}\rm{C}$ as a
	 function of photon energy at a fixed laboratory 
 	 angle of $\theta_{\rm lab}=60^{\circ}$, with and
	 without pionic distortions. Tensor (dashed lines) 
	 and vector (solid lines) parameterizations of the 
	 elementary amplitude are used. The experimental 
	 data is from Ref.~\protect\cite{schmitz96}.} 
\label{fig8}
\end{figure}

\begin{figure}[h]
 \null
 \vskip1.2in
 \includegraphics{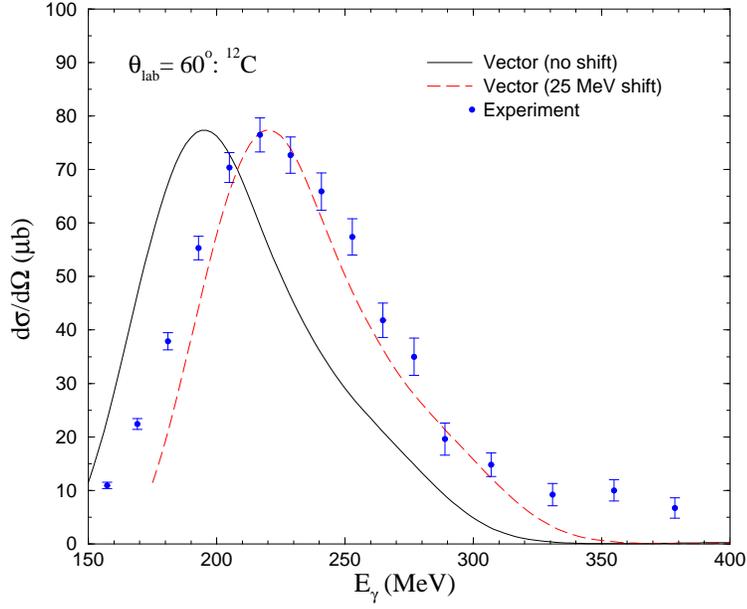}
 \vskip2.0in
\caption{Differential cross section for the coherent pion
	 photoproduction reaction from $^{12}\rm{C}$ as a
	 function of photon energy at a fixed laboratory 
 	 angle of $\theta_{\rm lab}=60^{\circ}$, with
	 pionic distortions and using only a vector 
	 parameterization of the elementary amplitude.
 	 The same calculation---including a shift of 25 MeV 
	 is also included (dashed line). The experimental 
	 data is from Ref.~\protect\cite{schmitz96}.}
\label{fig9}
\end{figure}

We conclude this section by presenting in Figs.~\ref{fig8} and
\ref{fig9}, a comparison between our plane- and distorted-wave
calculations with experimental data for the coherent cross section 
from $^{12}$C as a function of photon energy for a fixed angle of 
$\theta_{\rm lab}\!=\!60^{\circ}$. The experimental data from
MAMI is contained in the doctoral dissertation of
M. Schmitz~\cite{schmitz96}. 

Perhaps the most interesting feature in these figures is the very good
agreement between our RDWIA calculation using the vector
representation and the data---if we were to shift our results by
+25~MeV. Indeed, this effect is most clearly appreciated in
Fig.~\ref{fig9} where the shifted calculation is now represented by
the dashed line. In our treatment of the coherent process, the
detailed shape of the cross section as a function of energy results
from a delicate interplay between several effects arising from: a) the
elementary amplitude---which peaks at the position of the delta
resonance ($E_{\gamma}\simeq 340$~MeV from a free nucleon and slightly
lower here because of the optimal prescription~\cite{pisabe97}), b) the
nuclear form factor---which peaks at low-momentum transfer, and c) the
pionic distortions---which strongly quench the cross sections at high
energy, as more open channels become available. We believe that the
pionic distortions (see Sec.~\ref{sec:distor}) as well as the nuclear
form factor have been modeled accurately in the present work. The
elementary amplitude, although obtained from a recent phase-shift
analysis by the VPI group~\cite{Arnphn}, remains one of the biggest
uncertainties, as no microscopic model has been used to estimate
possible medium modifications to the on-shell amplitude.  Evidently,
an important modification might arises from the production,
propagation, and decay of the $\Delta$-resonance in the nuclear
medium. Indeed, a very general result from hadronic physics, obtained
from analyses of quasielastic $(p,n)$ and $({}^{3}He,t)$
experiments~\cite{Gaarde91}, is that the position of the $\Delta$-peak
in nuclear targets is lower relative to the one observed from a free
proton target.
 
However, it is also well known that such a shift is not observed when
the $\Delta$-resonance is excited electromagnetically~\cite{Gaarde91}.
This apparent discrepancy has been attributed to the different dynamic
responses that are being probed by the two processes. In the case of
the hadronic process, it is the (pion-like) spin-longitudinal response
that is being probed, which is known to get ``softened'' (shifted to
lower excitation energies) in the nuclear medium. Instead,
quasielastic electron scattering probes the spin-transverse
response---which shows no significance energy shift.  Unfortunately,
in our present local-impulse-approximation treatment it becomes
impossible to assess the effects associated with medium modifications
to the $\Delta$-resonance. A detailed study of possible violations to
the impulse approximation and to the local assumption remains an
important open problem for the future (for a qualitative discussion
see Sec.~\ref{sec:impviol}).

\section{Violations to the Impulse Approximation}
\label{sec:impviol}
\begin{figure}[h]
 \null
 \vskip1.2in
 \includegraphics{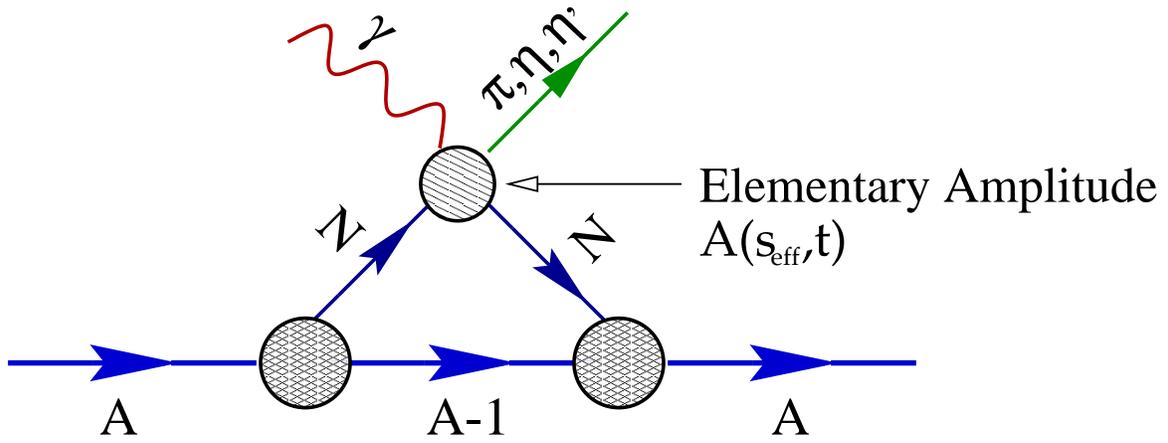}
 \vskip1.8in
\caption{Pictorial representation of the impulse
	 approximation for the coherent photoproduction of 
	 pseudoscalar mesons. Note that the elementary amplitude 
	 is evaluated using the optimal prescription (see, for
	 example, Ref~\protect\cite{pisabe97}.)}
\label{fig10}
\end{figure}

In this section we address an additional ambiguity in the formalism,
namely, the use of the impulse approximation. The basic assumption
behind the impulse approximation is that the interaction in the medium
is unchanged relative to its free-space value. The immense
simplification that is achieved with this assumption is that the
elementary interaction now becomes model independent, as it can be
obtained directly from a phase-shift analysis of the experimental data
(see, for example, Ref.~\cite{Arnphn}). The sole remaining question 
to be answered is the value of $s$ at which the elementary amplitude 
should be evaluated, as now the target nucleon is not free but
rather bound to the nucleus (see Fig.~\ref{fig10}). This question 
is resolved by using the ``optimal'' prescription of Gurvitz, 
Dedonder, and Amado~\cite{GDA79}, which suggests that the elementary 
amplitude should be evaluated in the Breit frame. Then, this optimal 
form of the impulse approximation leads to a factorizable and local 
scattering amplitude---with the nuclear-structure information 
contained in a well-determined vector form factor. Moreover, as the 
final-state interaction between the outgoing meson and the nucleus 
is well constrained from other data, a parameter-free calculation 
of the coherent photoproduction process ensues.

\begin{figure}[h]
 \null
 \vskip2.0in
 \includegraphics{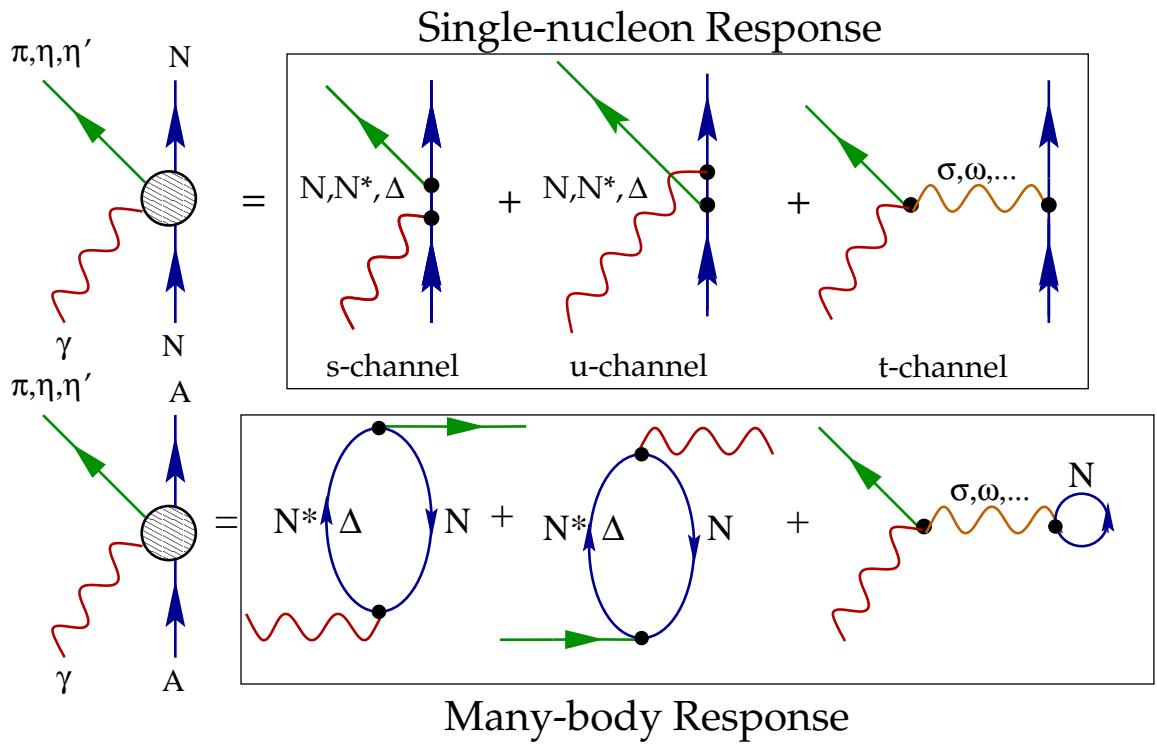}
 \vskip2.2in
\caption{Characteristic s-, u-, and t-channel Feynman 
	 diagrams for the photoproduction of pseudoscalar
	 mesons from a single nucleon (upper panel) 
	 and---coherently---from the nucleus (lower panel).}
\label{fig11}
\end{figure}

This form of the impulse approximation has been used with great
success in hadronic processes, such as in $(p,p')$ and $(p,n)$
reactions, and in electromagnetic processes, such as in electron
scattering. Perhaps the main reason behind this success is that the
elementary nucleon-nucleon or electron-nucleon interaction is mediated
exclusively by $t$-channel exchanges---such as arising from 
$\gamma$-, $\pi$-, or $\sigma$-exchange. This implies that the local 
approximation (i.e., the assumption that the nuclear-structure 
information appears exclusively in the form of a local nuclear form 
factor) is well justified. For the coherent process this would also 
be the case if the elementary amplitude would be dominated by the 
exchange of mesons, as in the last Feynman diagram in
Fig.~\ref{fig11}. However, it is well known---at least for the 
kinematical region of current interest---that the elementary
photoproduction process is dominated by resonance ($N^{\star}$ or
$\Delta$) formation, as in the $s$-channel Feynman diagram of
Fig.~\ref{fig11}. This suggests that the coherent reaction probes, in
addition to the nuclear density, the polarization structure of the
nucleus (depicted by the ``bubbles'' in Fig.~\ref{fig11}). As the
polarization structure of the nucleus is sensitive to the ground- as
well as to the excited-state properties of the nucleus, its proper
inclusion could lead to important corrections to the local
impulse-approximation treatment. Indeed, Peters, Lenske, and Mosel
have lifted the local assumption and have reported---in contrast to
all earlier local studies---that the $S_{11}(1535)$ resonance does
contribute to the coherent photoproduction of $\eta$-mesons. Clearly,
understanding these additional contributions to the coherent process
is an important area for future work.

\section{Conclusions}
\label{sec:concl}

We have studied the coherent photoproduction of pseudoscalar
mesons in a relativistic-impulse-approximation approach. We 
have placed special emphasis on the ambiguities underlying 
most of the current theoretical approaches. Although our 
conclusions are of a general nature, we have focused our 
discussions on the photoproduction of neutral pions due
to the ``abundance'' of data relative to the other 
pseudoscalar channels.

We have employed a relativistic formalism for the elementary 
amplitude as well as for the nuclear structure. We believe 
that, as current relativistic models of nuclear structure 
rival some of the most sophisticated nonrelativistic ones, 
there is no longer a need to resort to a nonrelativistic 
reduction of the elementary amplitude. Rather, the full 
relativistic structure of the coherent amplitude should 
be maintained~\cite{pisabe97,aps98}.

We have also extended our treatment of the pion-nucleus 
interaction to the $\Delta$-resonance region. Although 
most of the details about the optical potential will be 
reported shortly~\cite{raddad98}, we summarize briefly
some of our most important findings. As expected, pionic 
distortions are of paramount importance. Indeed, we have 
found factors-of-two enhancements (at low energies) and 
up to factors-of-five reductions (at high energies) in 
the coherent cross section relative to the plane-wave 
values. Yet, ambiguities arising from the various choices 
of optical-model parameters are relatively small; of at 
most 30\%.

By far the largest uncertainty in our results emerges from 
the ambiguity in extending the many---actually 
infinite---equivalent representations of the elementary 
amplitude off the mass shell. While all these choices are 
guaranteed to give identical results for on-shell observables, 
they yield vastly different predictions off-shell. In this 
work we have investigated two such representations: a tensor 
and a vector. The tensor representation employs the ``standard'' 
form of the elementary 
amplitude~\cite{bentan90,cgln57} and generates a coherent 
photoproduction amplitude that is proportional to the isoscalar 
tensor density. However, this form of the elementary amplitude, 
although standard, is not unique. Indeed, through a simple 
manipulation of operators between on-shell Dirac spinors, the 
tensor representation can be transformed into the vector
one, so-labeled because the resulting coherent amplitude becomes 
proportional now to the isoscalar vector density. The tensor and 
vector densities were computed in a self-consistent, mean-field 
approximation to the Walecka model~\cite{serwal86}. The Walecka 
model is characterized by the existence of large Lorentz scalar 
and vector potentials that are responsible for a large
enhancement of the lower components of the single-particle wave 
functions. This so-called ``$M^{\star}$-enhancement'' generates 
a large increase in the tensor density, as compared to a
scheme in which the lower component is computed from the
free-space relation. No such enhancement is observed in
the vector representation, as the vector density is 
insensitive to the $M^{\star}$-effect. As a result, the tensor 
calculation predicts coherent photoproduction cross sections 
that are up to a factor-of-five larger than the vector results. 
These large enhancement factors are not consistent with 
existent experimental data. Still, it is important to note that 
the vastly different predictions of the two models have been
obtained using the same pionic distortions, the same 
nuclear-structure model, and two sets of elementary amplitudes 
that are identical on-shell.

Finally, we addressed---in a qualitative fashion---violations to the
impulse approximation. In the impulse approximation one assumes that
the elementary amplitude may be used without modification in the
nuclear medium. Moreover, by adopting the optimal prescription of
Ref.~\cite{GDA79}, one arrives at a form for the coherent amplitude
that is local and factorizable.  Indeed, such an optimal form has been
used extensively---and with considerable success---in electron and
nucleon elastic scattering from nuclei. We suggested here that the
reason behind such a success is the $t$-channel--dominance of these
processes. In contrast, the coherent-photoproduction process is
dominated by resonance formation in the $s$-channel. In the nuclear
medium a variety of processes may affect the formation, propagation,
and decay of these resonances. Thus, resonant-dominated processes may
not be amenable to treatment via the impulse-approximation. Further,
in $s$-channel--dominated processes, it is not the local nuclear
density that is probed, but rather, it is the (non-local) polarization
structure of the nucleus.  This can lead to important deviations from
the naive local picture.  Indeed, by relaxing the local assumption,
Peters and collaborators have reported a non-negligible contribution
from the $S_{11}(1535)$ resonance to the coherent photoproduction of
$\eta$-mesons~\cite{Peters98}, in contrast to all earlier local
studies.

In summary, we have studied a variety of sources that challenge
earlier studies of the coherent photoproduction of pseudoscalar 
mesons. Without a clear understanding of these issues, erroneous 
conclusions are likely to be extracted from the wealth of 
experimental data that will soon become available. What will be
the impact of these calculations on our earlier work on the coherent 
photoproduction of $\eta$-mesons~\cite{pisabe97,aps98} is hard to 
predict. Yet, based on our present study it is plausible that the 
large enhancement predicted by the tensor form of the elementary 
amplitude might not be consistent with the experimental data.
In that case, additional calculations using the vector form will 
have to be reported. Moreover, this should be done within a 
framework that copes simultaneously with all other theoretical
ambiguities. Indeed, many challenging and interesting lessons 
have yet to be learned before a deep understanding of the 
coherent-photoproduction process will emerge.

\acknowledgments
We are indebted to J.A. Carr for many useful discussions on the 
topic of the pion-nucleus optical potential, and to R. Beck and
B. Krusche for many conversations on experimental issues. One of 
us (LJA) thanks  W.~Peters for many illuminating (electronic) 
conversations. This work was supported in part by the U.S. 
Department of Energy under Contracts Nos. DE-FC05-85ER250000 
(JP), DE-FG05-92ER40750 (JP) and by the U.S. National Science 
Foundation (AJS).

\appendix
\section{Pion-Nucleus Optical Potential}
\label{sec:appendix}

        The form of the optical potential is derived using a
        semi-phenomenological formalism that uses a parameterized form
        of the elementary $\pi N\!\rightarrow\!\pi N$ amplitude that is
        assumed to remain unchanged in the nuclear medium (impulse
        approximation). However, the elementary amplitude does not
        encompass the many other processes that can occur in the
        many-body environment, such as multiple scattering, true pion
        absorption, Pauli blocking, and Coulomb (in the case of
        charged-pion scattering) interactions. The corrections
        resulting from these processes are of second and higher 
	order relative to the strength of the first-order
        expression given by the impulse approximation. To account for
        these corrections, the impulse approximation form of the
        optical potential is modified to arrive at a pion-nucleus
        optical potential---applicable from threshold up to the
        delta-resonance region---of the form:
\begin{eqnarray}
        2 \omega U &\!=\!&\!-4\pi 
        \left[
           p_1 b(r) \!+\! p_2 B(r)\!-\! 
           \vec{\nabla} Q(r) \cdot\vec{\nabla}\!-\! 
           \frac {1}{4} p_1 u_1 {\nabla}^2 c(r)\!-\! 
           \frac {1}{4} p_2 u_2 {\nabla}^2 C(r)\!+\! 
           p_1 y_1 \widetilde{K}(r)
        \right]\;, 
\end{eqnarray} 
where 
\begin{mathletters}
 \begin{eqnarray}
   b(r) &=& \bar{b}_0\rho(r)-\epsilon_\pi {b_1} {\delta}\rho(r) \;, \\ 
   B(r) &=& {B_0}\rho^2(r)-\epsilon_\pi {B_1}\rho(r)\delta\rho(r) \;, \\ 
   c(r) &=&  c_0\rho(r)-\epsilon_\pi {c_1}{\delta}\rho(r)\;, \\ 
   C(r) &=&  C_0\rho^2(r)-\epsilon_\pi{C_1}\rho(r){\delta}\rho(r)\;, \\ 
   Q(r) &=& \frac{ L(r) }{1 + \frac {4 \pi}{3} \lambda L(r)} + p_1 x_1
            \acute{c} \rho(r)\;, \\
   L(r) &=& p_1 x_1 c(r) + p_2 x_2 C(r) \;, \\  
   \widetilde{K}(r) &=& \frac {3}{5} \left({\frac{3
   \pi^2}{2}}\right)^{2/3} c_0\rho^{5/3}(r)\;,    
 \end{eqnarray}
\end{mathletters}
and with
\begin{mathletters}
 \begin{eqnarray}
   \bar{b}_0 &=& {b_0} - p_1 \frac {A-1}{A} (b^2_0 + 2 b^2_1) I \;, \\ 
   \acute{c} &=& p_1 x_1 \frac {1}{3} k^2_o (c^2_0 + 2 c^2_1) I \;. 
 \end{eqnarray}
\end{mathletters}
In the above expressions, the set \{$p_1, u_1, x_1,$ and $y_1$\}
represents various kinematic factors in the effective $\pi-N$ system
(pion-nucleon mechanisms), and the set \{$p_2, u_2,$ and $x_2$\}
represents the corresponding kinematic factors in the $\pi\!-\!2N$
system (pion-two-nucleon mechanisms). These kinematic factors have
been derived using the relativistic potential model~\cite{KeMcTh59}
with no recourse to nonrelativistic approximations and it includes
nucleus recoil. The set of parameters \{$b_0, b_1, c_0,$ and $c_1$\}
originates from the $\pi N\!\rightarrow\!\pi N$ elementary amplitudes
while all other parameters--excluding the kinematic factors--have
their origin in the second and higher order corrections to the optical
potential. These first-order parameters have been determined from a
recent $\pi\!-\!N$ phase-shift analysis~\cite{Arnpin}, in contrast to
the approach by Carr and collaborators in which they were fit to
pionic-atom results.  In spite of this difference, the parameters
determined by the two methods match nicely. Nuclear effects enter in
the optical potentials through the nuclear density $\rho(r)$, and
through the neutron-proton density difference
$\delta\rho(r)$. Moreover, $A$ is the atomic number, $\lambda$ is the
Ericson-Ericson effect parameter, $k_o$ is the pion lab momentum,
$\omega$ is the pion energy in the pion-nucleus center of mass system,
and $I$ is the so-called $1/r_{correlation}$ function. The $B$ and $C$
parameters arise from true pion absorption. A detailed account of this
optical potential will be the subject of a paper that will be
submitted for publication shortly~\cite{raddad98}.


\vfill\eject
%
\vfill\eject
\vfill\eject
\vfill\eject
\vfill\eject
\vfill\eject	
\vfill\eject
\vfill\eject
\vfill\eject
\vfill\eject
\vfill\eject
\vfill\eject
%
\mediumtext
 \begin{table}
  \caption{Maxima of the differential cross section 
	   (in $\mu$b) for the coherent pion
	   photoproduction reaction from ${}^{12}$C
	   at various energies.}
   \begin{tabular}{cccc}
   $E_{\gamma}$~(MeV) & Tensor & Vector& Experiment    \\
   \tableline
      $235$  &  $694$  &  $116$  &  $105$  \\  
      $250$  &  $731$  &  $133$  &  $190$  \\    
      $291$  &  $786$  &  $186$  &  $175$  \\
   \end{tabular}
  \label{table1}
 \end{table}
\vfill\eject
%
\end{document}